\documentclass[preprint,prl,endfloats*,superscriptaddress]{revtex4-2}

\usepackage{graphicx}
\usepackage{dcolumn}
\usepackage{bm}
\usepackage{color}
\usepackage{amsmath}
\usepackage{latexsym}
\usepackage{hyperref} 
\usepackage{pifont}
\usepackage{multirow}

\newcommand{\rfig}[1]{Fig.~\ref{#1}}

\newcommand{\req}[1]{Eq.~(\ref{#1})}

\makeindex

\begin{document}

\title{Orbital Magneto-Nonlinear Anomalous Hall Effect in Kagome Magnet Fe$_3$Sn$_2$}

\author{Lujunyu Wang}
\thanks{These authors contributed equally.}
\affiliation{State Key Laboratory for Artificial Microstructure and Mesoscopic Physics, Frontiers Science Center for Nano-optoelectronics, Peking University, Beijing 100871, China}
\author{Jiaojiao Zhu}
\thanks{These authors contributed equally.}
\affiliation{Research Laboratory for Quantum Materials, Singapore University of Technology and Design, Singapore 487372, Singapore}
\author{Haiyun Chen}
\thanks{These authors contributed equally.}
\affiliation{State Key Laboratory for Artificial Microstructure and Mesoscopic Physics, Frontiers Science Center for Nano-optoelectronics, Peking University, Beijing 100871, China}
\author{Hui Wang}
\thanks{These authors contributed equally.}
\affiliation{Research Laboratory for Quantum Materials, Singapore University of Technology and Design, Singapore 487372, Singapore}
\affiliation{Division of Physics and Applied Physics, School of Physical and Mathematical Sciences, Nanyang Technological University, Singapore 637371, Singapore}
\author{Jinjin Liu}
\affiliation{Centre for Quantum Physics, Key Laboratory of Advanced Optoelectronic Quantum Architecture and Measurement (MOE), School of Physics, Beijing Institute of Technology, Beijing 100081, China}
\author{Yue-Xin Huang}
\affiliation{Research Laboratory for Quantum Materials, Singapore University of Technology and Design, Singapore 487372, Singapore}
\affiliation{School of Sciences, Great Bay University, Dongguan 523000, China}
\author{Bingyan Jiang}
\author{Jiaji Zhao}
\author{Hengjie Shi}
\author{Guang Tian}
\author{Haoyu Wang}
\affiliation{State Key Laboratory for Artificial Microstructure and Mesoscopic Physics, Frontiers Science Center for Nano-optoelectronics, Peking University, Beijing 100871, China}
\author{Yugui Yao}
\affiliation{Centre for Quantum Physics, Key Laboratory of Advanced Optoelectronic Quantum Architecture and Measurement (MOE), School of Physics, Beijing Institute of Technology, Beijing 100081, China}
\affiliation{Beijing Key Lab of Nanophotonics and Ultrafine Optoelectronic Systems, Beijing Institute of Technology, Beijing 100081, China}
\affiliation{Material Science Center, Yangtze Delta Region Academy of Beijing Institute of Technology, Jiaxing, China}
\author{Dapeng Yu}
\affiliation{Shenzhen Institute for Quantum Science and Engineering, Southern University of Science and Technology, Shenzhen 518055, China}
\author{Zhiwei Wang}
\email{zhiweiwang@bit.edu.cn}
\affiliation{Centre for Quantum Physics, Key Laboratory of Advanced Optoelectronic Quantum Architecture and Measurement (MOE), School of Physics, Beijing Institute of Technology, Beijing 100081, China}
\affiliation{Beijing Key Lab of Nanophotonics and Ultrafine Optoelectronic Systems, Beijing Institute of Technology, Beijing 100081, China}
\affiliation{Material Science Center, Yangtze Delta Region Academy of Beijing Institute of Technology, Jiaxing, China}
\author{Cong Xiao}
\email{congxiao@um.edu.mo}
\affiliation{Institute of Applied Physics and Materials Engineering, University of Macau, Taipa, Macau, China}
\affiliation{Department of Physics, The University of Hong Kong, Hong Kong, China}
\affiliation{HKU-UCAS Joint Institute of Theoretical and Computational Physics at Hong Kong, China}
\author{Shengyuan A. Yang}
\affiliation{Institute of Applied Physics and Materials Engineering, University of Macau, Taipa, Macau, China}
\author{Xiaosong Wu}
\email{xswu@pku.edu.cn}
\affiliation{State Key Laboratory for Artificial Microstructure and Mesoscopic Physics, Frontiers Science Center for Nano-optoelectronics, Peking University, Beijing 100871, China}
\affiliation{Shenzhen Institute for Quantum Science and Engineering, Southern University of Science and Technology, Shenzhen 518055, China}
\affiliation{Collaborative Innovation Center of Quantum Matter, Beijing 100871, China}
\affiliation{Peking University Yangtze Delta Institute of Optoelectronics, Nantong 226010, Jiangsu, China}

\begin{abstract}
It has been theoretically predicted that perturbation of the Berry curvature by electromagnetic fields gives rise to intrinsic nonlinear anomalous Hall effects that are independent of scattering. Two types of nonlinear anomalous Hall effects are expected. The electric nonlinear Hall effect has recently begun to receive attention, while very few studies are concerned with the magneto-nonlinear Hall effect. Here, we combine experiment and first-principles calculations to show that the kagome ferromagnet Fe$_3$Sn$_2$ displays such a magneto-nonlinear Hall effect. By systematic field angular and temperature-dependent transport measurements, we unambiguously identify a large anomalous Hall current that is linear in both applied in-plane electric and magnetic fields, utilizing a unique in-plane configuration. We clarify its dominant orbital origin and connect it to the magneto-nonlinear Hall effect. The effect is governed by the intrinsic quantum geometric properties of Bloch electrons. Our results demonstrate the significance of the quantum geometry of electron wave functions from the orbital degree of freedom and open up a new direction in Hall transport effects.
\end{abstract}



\maketitle

Inspired by the understanding of the Berry physics and topological materials, it is now well accepted that the properties of quantum materials are determined by not only the band dispersion but also the geometry of the quantum wave functions, such as the Berry curvature and the quantum metric. Consequently, probing and controlling the quantum geometry is becoming one of the main themes in the field of quantum materials~\cite{Xu2018Sep,Lai2021Aug,Ye2023Jan,Ma2021Dec,Sinha2022Jul}. The nonlinear anomalous Hall effect (NLHE) has recently attracted enormous interest~\cite{Sodemann2015Nov,Facio2018Dec,Du2018Dec,Ma2019,Kang2019,Xiao2019Oct,Matsyshyn2019Dec,Kumar2021Jan,Du2021Aug,Du2021Auga,Ma2021Dec,Sinha2022Jul,Ma2022Sep}. It probes the so-called Berry curvature dipole. The Berry curvature is usually taken as field-independent in this effect. On the other hand, it has been predicted early on that electromagnetic fields can influence the Berry curvature, giving rise to a different NLHE \cite{Gao2014Apr}. The effect is intrinsic as it is determined by the geometric property of wave functions and independent of scattering. Since both electric and magnetic fields can modify the Berry curvature, two types of intrinsic effects are expected, i.e., electric NLHE (eNLHE) and magneto-NLHE (mNLHE). The eNLHE is closely related to the quantum metric that defines the infinitesimal distance in the Hilbert space on the Brillouin zone. The mNLHE is associated with the Christoffel symbol that defines the affine geometry of the Brillouin zone. These two geometric properties together make the Brillouin zone a Riemannian manifold~\cite{Gao2014Apr}. Therefore, studies of two NLHEs provide complementary information on the quantum geometry.

Detection of the eNLHE using harmonic measurements is relatively straightforward, as has been done for the NLHE due to the field-independent Berry curvature dipole~\cite{Sodemann2015Nov,Ma2019,Kang2019}. To distinguish the eNLHE from the NLHE due to the field-independent Berry curvature dipole, proposals have been put forth and particular attention has been paid to $\mathcal{PT}$-symmetric materials, in which the latter is precluded by symmetry~\cite{Wang2021Dec,Liu2021Dec,Kirikoshi2023Apr}. Experimental studies have appeared very recently \cite{Gao2023Jun,Wang2023Sep}. In contrast, identification of the mNLHE seems very tricky, because the effect, scaling as $EH$, shares the same electromagnetic field dependence with the ordinary Hall effect due to the Lorentz force~\cite{Gao2014Apr}. Here, $E$ and $H$ are the electric and magnetic fields, respectively. Although the symmetry requirement for the mNLHE is less stringent than the eNLHE, which needs inversion symmetry breaking, studies on the mNLHE are scarce. A recent discovery of a novel in-plane anomalous Hall effect (IAHE) on heterodimensional superlattice V$_5$S$_8$ provides a unique opportunity for probing the mNLHE~\cite{Zhou2022Sep}, owing to a strong suppression of the ordinary Hall effect in the special in-plane configuration. In this work, we study the the Hall effect in a kagome ferromagnetic semimetal Fe$_3$Sn$_2$ and observe a strong mNLHE. The ferromagnetic order of the material enables us to identify the dominance of the orbital contribution. The first-principles calculations are in excellent agreement with the experiment data.

To understand the origin of this orbital mNLHE, we recall that for the usual anomalous Hall effect in magnets, there is an important contribution from so-called anomalous velocity $\sim \boldsymbol{E} \times \bm \Omega$ of Bloch electrons, where  $\bm \Omega$ is the Berry curvature of band structure~\cite{Nagaosa2010,Xiao2010}. A magnetic field perturbs the band structure, therefore giving a correction to the Berry curvature~\cite{Gao2014Apr}. In the extended semiclassical theory~\cite{Wang2024Jan},
the $H$-field-induced Berry connection is given by
\begin{equation}
\mathcal{A}_{b}^{H}\left( \boldsymbol{k}\right) = \mu_{0}H_{a}[F_{ab}^{\text{S}}\left(  \boldsymbol{k}\right)  +F_{ab}^{\text{O}}\left( \boldsymbol{k}\right) ].\label{F}%
\end{equation}
Here the subscripts $a$ and $b$ denote Cartesian components, $\boldsymbol{k}$ is the wave vector, and we suppress the band indices. It is important to note that both $\mathcal{A}^{H}$ and $F$'s are gauge invariant.
The coefficients $F_{ab}^{\text{S}}$ and $F_{ab}^{\text{O}}$ are known as anomalous spin polarizability and anomalous orbital polarizability (AOP), respectively. They represent the susceptibility of positional shift of Bloch wave packets with respect to an applied $H$ field due to its coupling with spin and orbital degrees of freedom. The induced Berry connection generates the correction to the Berry curvature
$\boldsymbol{\Omega}^{H}=\nabla_{\boldsymbol{k}}\times\boldsymbol{\mathcal{A}}^{H}$, which in turn leads to the anomalous velocity and the anomalous Hall effect scaling as $\sim EH$. Clearly, in this simple picture, the orbital mechanism is associated with AOP tensor $F_{ab}^{\text{O}}$, which is an intrinsic band geometric quantity. The key point is that AOP is greatly enhanced and usually dominates over anomalous spin polarizability at bands near degeneracy regions. Thus, there is a chance that the orbital contribution can dominate over spin contribution in certain materials with entangled multiband structures around Fermi level, especially those topological band crossings. Nevertheless, such dominant orbital mNLHE has not been experimentally verified so far.

\begin{figure}[htbp]
	\begin{center}
		\includegraphics[width=1\columnwidth]{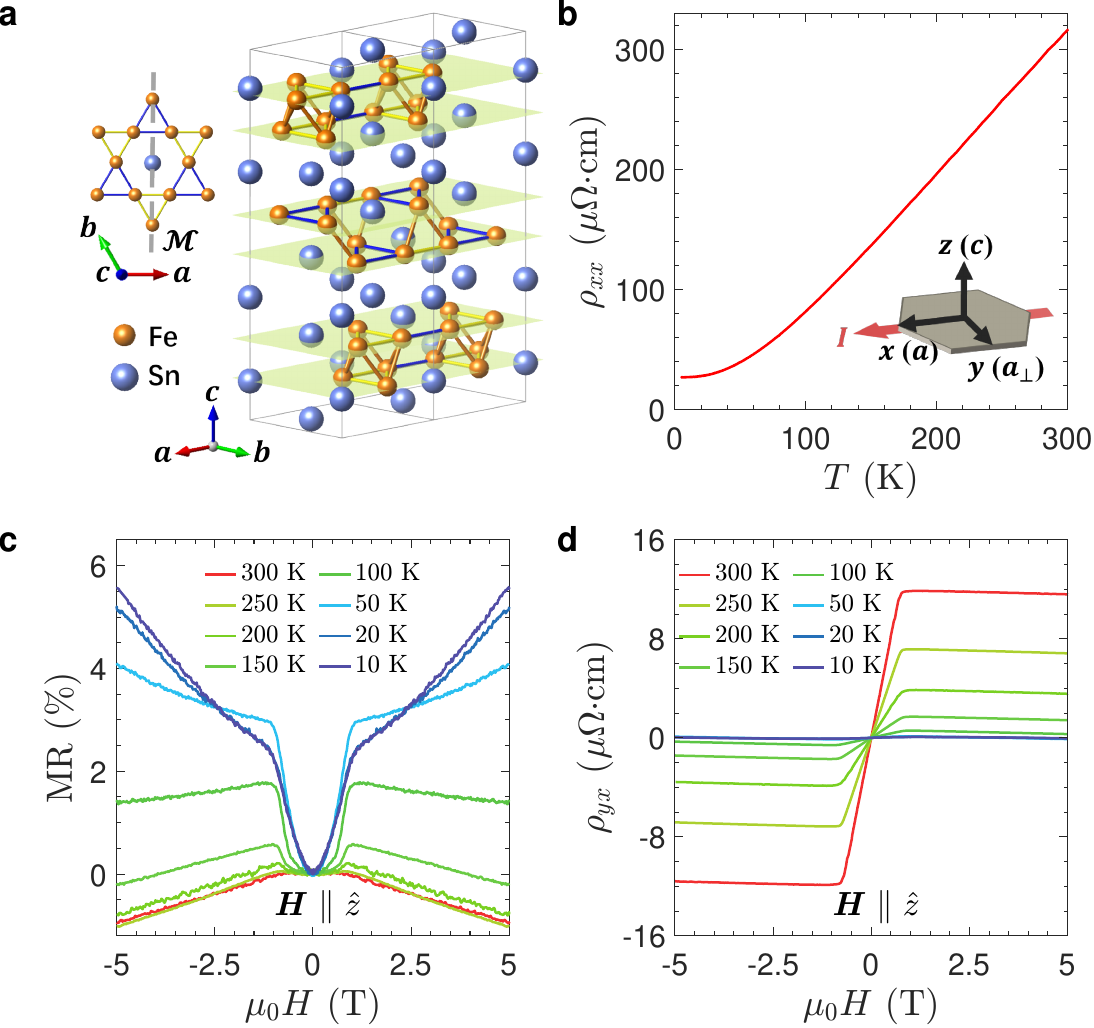}
		\caption{Crystal structure and basic transport characterization of Fe$_{3}$Sn$_{2}$. $\mathbf{a}$, Crystal structure. The crystal includes a honeycomb Sn layer sandwiched between kagome Fe$_{3}$Sn bilayers (green slices). The left illustration shows the breathing kagome Fe$_{3}$Sn layer.  The gray dashed line indicates the mirror plane perpendicular to the $a$ axis. $\mathbf{b}$, Temperature-dependent longitudinal resistivity along the $a$ axis. The inset depicts the current direction. The $x$, $y$, and $z$ axes correspond to the $a$, $a_{\perp}$, and $c$ axes, respectively. $\mathbf{c}$,$\mathbf{d}$, Magnetoresistance (${\text{MR}}=\frac{\rho_{xx}(H)-\rho_{xx}(0)}{\rho_{xx}(0)}$) and Hall resistivity under an out-of-plane field at various temperatures, respectively.}
		\label{Crystal}
	\end{center}
\end{figure}

Fe$_{3}$Sn$_{2}$ is a soft ferromagnetic semimetal with a high Curie temperature of 657~K \cite{LeCaer1978Feb,Giefers2006Sep}. Its structure consists of breathing kagome Fe$_{3}$Sn bilayers and Sn honeycomb layers stacked along the $c$ axis~\cite{Ye2018Mar}, as shown in \rfig{Crystal}a. This layered rhombohedral crystal structure belongs to the $R\overline{3}m$ space group (point group $D_{3d}$) with a mirror plane $M_x$ perpendicular to the $a$ axis (see the inset of \rfig{Crystal}b for the coordinate system we adopt). The $M_x$ mirror has a strong constraint on IAHE: IAHE must vanish identically if $H$ field (and/or magnetization) is parallel to $x$ (i.e., preserving $M_x$), and it is allowed for $H$ field along other in-plane directions (e.g., the $y$ direction).
Recent studies suggested the emergence of Weyl nodes when the magnetization is in-plane~\cite{Tanaka2020Apr,Biswas2020Aug,Fang2022Jan,Kumar2022Jul,Ren2022Nov}. These degeneracy points can strongly affect electronic response and transport properties. Consistently, experiments have shown the unusually strong responses to magnetic fields applied in $z$ and $y$ directions \cite{Yin2018Oct,Li2019Nov}. As mentioned above, these topological features are beneficial for enhancing the possible orbital mNLHE.

Our Fe$_{3}$Sn$_{2}$ crystals were grown by a two-step growth method. As the starting material, the pure phase of Fe$_{3}$Sn$_{2}$ polycrystals were obtained with solid-state reaction at 770~$^\circ$C. The Fe$_{3}$Sn$_{2}$ single crystals with shiny surfaces were grown from these polycrystals by the chemical vapor transport method with iodine as the transport agent, in a temperature gradient from 650~$^\circ$C to 720~$^\circ$C. The crystal structure and the chemical composition were confirmed by x-ray diffraction and energy dispersive spectroscopy, respectively~\cite{SM_mNLHE}~\nocite{Liang2018,Zeng2020Aug,Kresse1994May,Kresse1996Oct,Blochl1994Dec,Perdew1996Oct,Marzari1997Nov,Souza2001Dec,Mostofi2008May}. Single crystals of thin hexagonal plates with a lateral dimension of 200--1000~$\mu \text{m}$ and a thickness of 15--60~$\mu \text{m}$ were selected for transport measurements. Samples were mounted on a motorized rotation stage with a resolution better than $0.02^\circ$. A Hall sensor was placed on the sample holder for accurate determination of the rotation angle. The resistivity and Hall effect were measured by a four-probe method using lock-in amplifiers and extracted by a standard symmetrizing/antisymmetrizing procedure. The data presented in the main text are mainly from sample \#03, whereas the in-plane angular dependence was measured on sample \#04. Additional data and the data for sample \#01 are shown in the Supplemental Material~\cite{SM_mNLHE}~\nocite{Liang2018,Zeng2020Aug,Kresse1994May,Kresse1996Oct,Blochl1994Dec,Perdew1996Oct,Marzari1997Nov,Souza2001Dec,Mostofi2008May}. These samples show basically identical behaviors.

The resistivity $\rho_{xx}$ of our crystal as a function of temperature is shown in \rfig{Crystal}b. $\rho_{xx}$ exhibits a typical metallic behavior, with the residual resistivity ratio $\rho_{xx}(300\ \text{K})/\rho_{xx}(5\ \text K)$ being about 12. We first apply external $H$ field in the $z$ direction. The magnetoresistance and Hall resistivity of our samples display characteristics reported before~\cite{Kida2011Mar,Wang2016a,Ye2018Mar,Kumar2022Jul}. At high temperatures, the high-field magnetoresistance in \rfig{Crystal}c is negative and linear in $H$, consistent with electron-magnon scattering~\cite{Raquet2002Jul}. With decreasing temperature, a quadratic Lorentz magnetoresistance appears. The low-field valley in magnetoresistance reflects the reorientation of magnetization from the (in-plane) easy axis to the $c$ axis~\cite{Kumar2019Dec}. It increases with decreasing temperature, signifying the development of the spin reorientation transition, in which the easy axis rotates from the $z$ axis toward the $xy$ ($ab$) plane and eventually lies in the $xy$ plane below 70~K \cite{Kumar2019Dec,Biswas2020Aug,Wu2021Aug}. As seen in \rfig{Crystal}d, the jump of the Hall resistivity $\rho_{yx}$ at the coercive field corresponds to the anomalous Hall resistivity. $\rho_{yx}$ above the coercive field is contributed by the ordinary Hall effect of Lorentz-force origin. Above 100~K, the ordinary Hall effect is linear in magnetic field, suggesting the dominance of one type of carrier. At low temperatures, the field dependence of the ordinary Hall effect deviates from a straight line, indicating the involvement of multiple bands, as shown in Fig.~S3b~\cite{SM_mNLHE}~\nocite{Liang2018,Zeng2020Aug,Kresse1994May,Kresse1996Oct,Blochl1994Dec,Perdew1996Oct,Marzari1997Nov,Souza2001Dec,Mostofi2008May}.

\begin{figure}[htbp]
	\begin{center}
		\includegraphics[width=1\columnwidth]{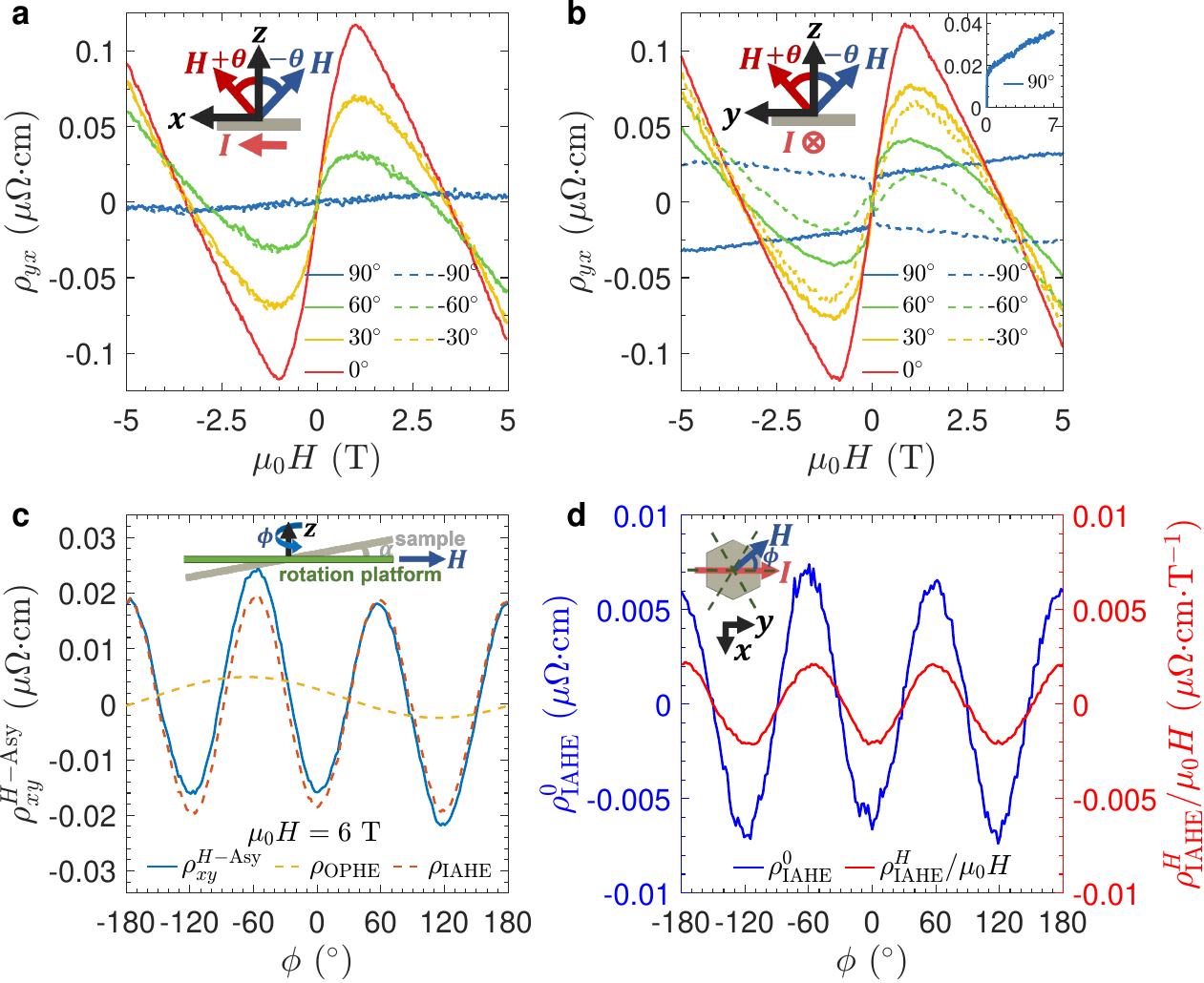}
		\caption{In-plane Hall effect and its angular dependence. $\mathbf{a}$, Hall resistivity for the magnetic field rotating in the $zx$ plane at $T=50$~K. $\mathbf{b}$, Hall resistivity for the magnetic field rotating in the $yz$ plane at $T=50$~K. The top-left insets in $\mathbf{a}$ and $\mathbf{b}$ depict the current and magnetic field directions. The top-right inset in $\mathbf{b}$ shows $\rho_\text{IAHE}$ from 0 to 7~T. $\mathbf{c}$, $H$-antisymmetric Hall resistivity $\rho_{xy}^{H\text{-Asy}}$ for magnetic field rotated in the $xy$ plane at $\mu_0 H=6$~T and $T=5$~K. $\rho_{xy}^{H\text{-Asy}}$ consists of two sinusoidal contributions with periods of $2\pi$ and $2/3\pi$. The $2\pi$ oscillation is contributed by the out-of-plane Hall component due to a misalignment sketched in the inset. $\mathbf{d}$, Angular dependence of $\rho_\text{IAHE}^{0}$ (blue, left axis) and $\rho_\text{IAHE}^H/\mu_{0}H$ (red, right axis) at $T=5$~K. The inset depicts the measurement configuration and mirrors (green dashed lines) of the crystal. }
		\label{RH}
	\end{center}
\end{figure}

To investigate the effect of magnetic field orientation on the Hall resistivity, we tilt the field in the $zx$ plane, as sketched in \rfig{RH}a. The angle between the field and the $z$ axis is denoted as $\theta$. $\rho_{yx}$ decreases with increasing angle and disappears when $H$ is in-plane. The behavior is consistent with the conventional picture of the ordinary (anomalous) Hall effect, in which the effect normally depends on the out-of-plane component of magnetic field (magnetization). This is corroborated by the observation that $\rho_{yx}$ at $+\theta$ coincides with that at $-\theta$. In stark contrast, $\rho_{yx}$ behaves distinctively when the field is rotated in the $yz$ plane. In particular, the Hall resistivities at $+\theta$ and $-\theta$ noticeably differ, implying that $\rho_{yx}$ is no longer solely determined by the out-of-plane components of field and magnetization. Furthermore, $\rho_{yx}$ remains substantial, even when the field lies in the $xy$ plane, indicating an in-plane Hall effect, as shown in \rfig{RH}b. Detailed measurements confirm that the observation is not due to an angle misalignment (see Supplemental Material~\cite{SM_mNLHE}~\nocite{Liang2018,Zeng2020Aug,Kresse1994May,Kresse1996Oct,Blochl1994Dec,Perdew1996Oct,Marzari1997Nov,Souza2001Dec,Mostofi2008May}). The in-plane Hall signal appears when the $H$ field is along the $y$ axis, while it vanishes when the field is along the $x$ axis. This behavior is consistent with the constraints imposed by the $M_x$ mirror we discussed above.

We then measure the angular dependence of the in-plane Hall effect for magnetic field rotated in the $xy$ plane. As shown in \rfig{RH}c, the Hall resistivity can be decomposed into two sinusoidal components. One has a period of $2\pi$, apparently resulting from the out-of-plane Hall effect due to a misalignment angle. The other dominant component is the in-plane Hall effect. It displays a period of $2\pi/3$, which results from the concerted action of the threefold rotational symmetry of the nonmagnetic lattice structure and the soft-magnet nature of Fe$_3$Sn$_2$ (detailed in the Supplemental Material~\cite{SM_mNLHE}~\nocite{Liang2018,Zeng2020Aug,Kresse1994May,Kresse1996Oct,Blochl1994Dec,Perdew1996Oct,Marzari1997Nov,Souza2001Dec,Mostofi2008May}). Here, we also remind that the in-plane Hall effect is odd in magnetic field~\cite{Liu2013Aug,Battilomo2021Jan,Cullen2021Jun,Sun2022Dec}, while the so-called planar Hall effect is even hence essentially an anisotropic magnetoresistance effect~\cite{Zhou2022Sep}. They can be separated by the antisymmetrizing or symmetrizing procedure.

Having confirmed the observation of IAHE, we further analyze its physical origin. As seen in the inset of \rfig{RH}b, the overall in-plane Hall signal apparently consists of two parts: a jump around zero field and a $H$-linear part away from the jump, i.e., $\rho_{yx}=\rho_\text{IAHE}^{0}+\rho_\text{IAHE}^H$. Both display an angular dependence with a period of $2\pi/3$, shown in \rfig{RH}d. Clearly, the jump is from the usual anomalous Hall effect due to magnetization (though in an unusual in-plane configuration), and this can be confirmed by its correlation with the field dependence of in-plane magnetization~\cite{Kumar2019Dec,Wu2021Aug}. The $H$-linear part $\rho_\text{IAHE}^H$ is consistent with the electromagnetic field dependence of the mNLHE. Since Fe$_3$Sn$_2$ is a soft magnet and the temperature is far below the Curie temperature, the magnetization of local moments should be well saturated after magnetic switching and therefore remains a constant~\cite{Kumar2019Dec}. Thus, the $H$-linear contribution can only be related to the spin and/or orbital coupling of carriers to the applied $H$ field. It is important to note that this contribution is remarkably large compared to  $\rho_\text{IAHE}^{0}$ due to magnetization. For example, at 10~K, $\rho_\text{IAHE}^H$ at 3~T already amounts to $\rho_\text{IAHE}^{0}$ (see Fig.~S3d in the Supplemental Material~\cite{SM_mNLHE}~\nocite{Liang2018,Zeng2020Aug,Kresse1994May,Kresse1996Oct,Blochl1994Dec,Perdew1996Oct,Marzari1997Nov,Souza2001Dec,Mostofi2008May}). As the exchange field of Fe$_3$Sn$_2$, which underlies $\rho_\text{IAHE}^{0}$, is on the order of 1000~T~\cite{Ye2018Mar}, one can see that the spin Zeeman coupling to applied $H$ field (a few Tesla) is negligible and cannot account for the large $\rho_\text{IAHE}^H$ that we observe. Meanwhile, AOP and its associated orbital contribution could be greatly enhanced in Fe$_3$Sn$_2$ due to the band crossing features around the Fermi level. All these imply that the pronounced IAHE is very likely to have a dominant orbital origin.

\begin{figure}[htbp]
	\begin{center}
		\includegraphics[width=1\columnwidth]{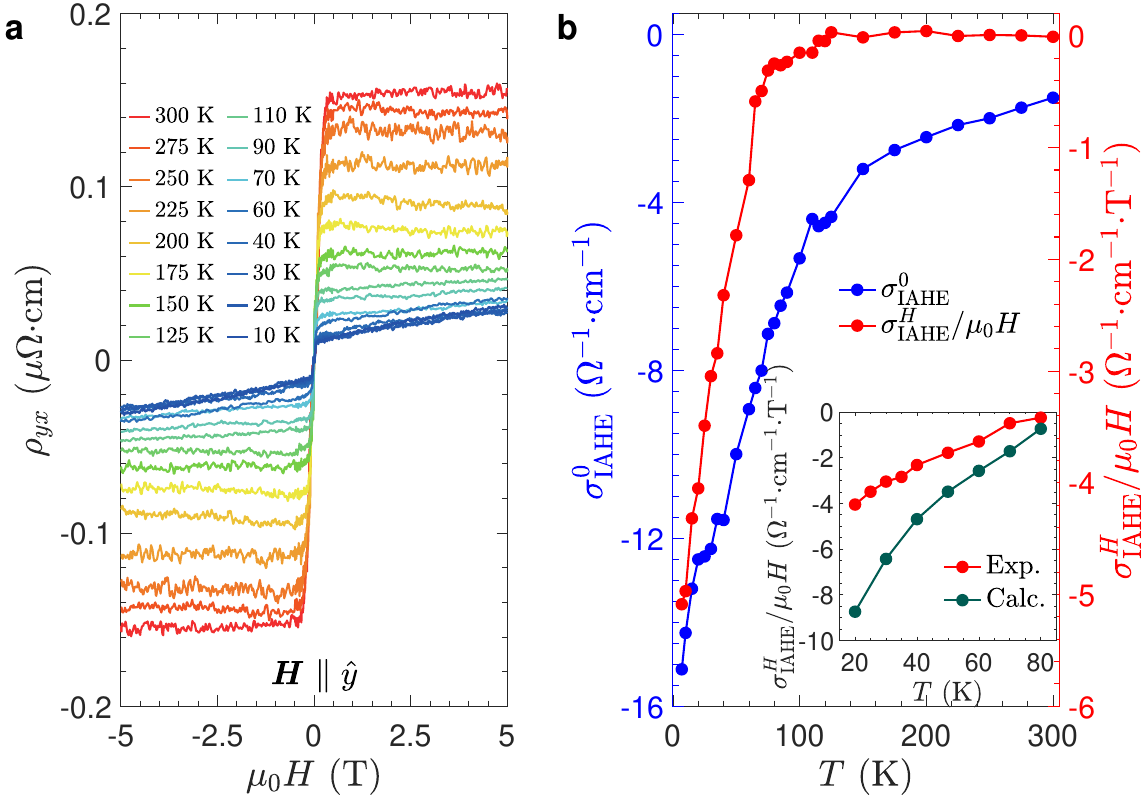}
		\caption{Temperature dependence of the in-plane Hall response. $\mathbf{a}$, Magnetic field dependence of the Hall resistivity for $H$ along the $y$ axis at various temperatures. The in-plane Hall signal consists of a jump around zero field and a $H$-linear dependence after that. $\mathbf{b}$, Hall conductivity $\sigma_\text{IAHE}^{0}$ (blue, left axis) and the slope $\sigma_\text{IAHE}^H/\mu_{0}H$ (red, right axis) for $H$ along the $y$ axis as functions of temperature. $\sigma_{\text {IAHE}}^{H}/\mu_{0}H$ from the experiment and the first-principles calculations are compared in the inset.}
		\label{RT}
	\end{center}
\end{figure}

We have also investigated the temperature dependence of the signal. \rfig{RT}a shows the total Hall resistivity $\rho_{yx}$ versus $H$ field for $\boldsymbol{H}\parallel\hat{y}$ at various temperatures from 10~K to 300~K. As temperature increases, $\rho_\text{IAHE}^{0}$ increases, while $\rho_\text{IAHE}^{H}$ decreases. By a linear fit to $\rho_{yx}$ above the jump, we extract $\rho_\text{IAHE}^{0}$ and $\rho_\text{IAHE}^{H}/\mu_0 H$ from the $y$-axis intercept and the slope, respectively. The corresponding conductivities can be calculated by $\sigma_\text{IAHE}^{0} \simeq -\frac{\rho_\text{IAHE}^{0}}{{\rho_{xx}}^2}$ and $\sigma_\text{IAHE}^H\simeq -\frac{\rho_\text{IAHE}^{H}}{{\rho_{xx}}^2}$, when $\rho_{yx} \ll \rho_{xx}$. \rfig{RT}b exhibits the temperature dependence of $\sigma_\text{IAHE}^{0}$ and $\sigma_{\text {IAHE}}^{H}/\mu_{0}H$.
$\sigma_\text{IAHE}^{0}$ strongly decreases with increasing temperature, although it persists to 300~K. $\sigma_{\text{IAHE}}^{H}/\mu_{0}H$ displays an even stronger suppression with increasing temperature and disappears above 120~K, shown in \rfig{RT}b. Such a dependence of $\sigma_{\text{IAHE}}^{H}$ is markedly different from the ordinary Hall effect (see Fig.~S4 in the Supplemental Material~\cite{SM_mNLHE}~\nocite{Liang2018,Zeng2020Aug,Kresse1994May,Kresse1996Oct,Blochl1994Dec,Perdew1996Oct,Marzari1997Nov,Souza2001Dec,Mostofi2008May}) and agrees well with the first-principles calculations, which will be discussed in the following.

\begin{figure}[htbp]
	\begin{center}
		\includegraphics[width=1\columnwidth]{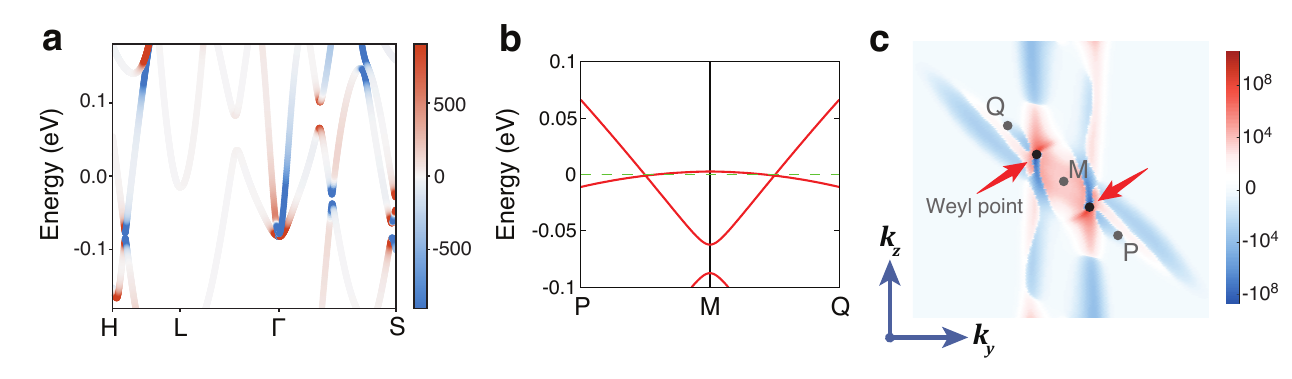}
		\caption{Calculated AOP of Fe$_{3}$Sn$_{2}$ with magnetization in the $y$ direction. $\mathbf{a}$, The band projection of $yy$ component of AOP in units of $\mu_{\text{B}}$\AA/eV, which is concentrated in small-gap regions. $\mathbf{b}$, Calculated band structures around two Weyl points appearing within an energy window of 1 meV about the Fermi surface of Fe$_{3}$Sn$_{2}$. The two Weyl points are located in the $k_x=0.793$ plane of the Brillouin zone. $\mathbf{c}$, Distribution of the $k$-resolved antisymmetrized AOP dipole on the Fermi surface, i.e., $f_{0}^{\prime}(v_y F_{yx}^\text{O}-v_x F_{yy}^\text{O})$, which is the integrand of Eq. (\ref{eq-Kai}) obtained via an integration by parts, in the $k_x=0.793$ plane of the momentum space. Here $f_{0}^{\prime}=\partial f_{0}/\partial \varepsilon$. The black points show a pair of Weyl points in this plane, which are the hot spots of AOP dipole. The coordinate of M point is (0.793, 0, 0).}
		\label{AOP}
	\end{center}
\end{figure}

To clarify the origin of this significant $H$-linear IAHE, we combine the theoretical formulation developed in Ref.~\onlinecite{Wang2024Jan} with first-principles calculations to yield quantitative estimations. Besides intrinsic contribution, our calculation also includes
extrinsic contributions quadratic in the relaxation time $\tau$ due to Lorentz force~\cite{Zhou2022Sep}.
The calculation details are presented in the Supplemental Material~\cite{SM_mNLHE}~\nocite{Liang2018,Zeng2020Aug,Kresse1994May,Kresse1996Oct,Blochl1994Dec,Perdew1996Oct,Marzari1997Nov,Souza2001Dec,Mostofi2008May}. Our calculation gives the nonlinear conductivity tensor $\chi$, defined by $j_{a}^\text{IAHE}=\chi_{abc}\mu_{0}E_{b}H_{c}$. It corresponds to the value $\sigma_{\text {IAHE}}^{H}/\mu_{0}H$ obtained in experiment. We find that the IAHE in Fe$_{3}$Sn$_{2}$ is indeed overwhelmingly dominated by the orbital contribution due to AOP, which is
\begin{equation}
\chi_{abc}^{\text{AOP}}=-\int[\text{d}\boldsymbol{k}]f_{0}\ (\partial_a F_{cb}^\text{O}-\partial_b F_{ca}^\text{O}), \label{eq-Kai}%
\end{equation}
where the integration with Fermi distribution $f_0$ is over all occupied states, and the derivative is with respect to the wave vector component. By an integration by parts, \req{eq-Kai} can also be put in the form of an integration of AOP over the Fermi surface. This contribution quantifies the antisymmetrized combination of AOP dipole, which is defined in parallel to the Berry curvature dipole \cite{Sodemann2015Nov} and quantum metric dipole \cite{Gao2023Jun,Wang2023Sep}. Our calculation shows that in the low-temperature regime ($<80$ K) this AOP contribution is 2 to 3 orders of magnitude larger than the spin contribution, and at least 1 order of magnitude larger than other contributions (Table~S1 in the Supplemental Material~\cite{SM_mNLHE}~\nocite{Liang2018,Zeng2020Aug,Kresse1994May,Kresse1996Oct,Blochl1994Dec,Perdew1996Oct,Marzari1997Nov,Souza2001Dec,Mostofi2008May}). This confirms the dominant orbital origin of IAHE in Fe$_{3}$Sn$_{2}$.

To correlate the large AOP contribution with band features in Fe$_{3}$Sn$_{2}$, in \rfig{AOP}a we plot the band projection of AOP ($F_{yy}^\text{O}$ component). One can see that it is indeed enhanced at small-gap regions. Taking a closer look at the band structure around Fermi surfaces of Fe$_{3}$Sn$_{2}$, we find four pairs of Weyl points within an energy window of 10 meV around the Fermi surface (see Table~S2 in the Supplemental Material~\cite{SM_mNLHE}~\nocite{Liang2018,Zeng2020Aug,Kresse1994May,Kresse1996Oct,Blochl1994Dec,Perdew1996Oct,Marzari1997Nov,Souza2001Dec,Mostofi2008May}), one of which connected by $M_x\mathcal{T}$ symmetry ($\mathcal{T}$ is the time reversal) are displayed in \rfig{AOP}b. These Weyl points act as hot spots generating large AOP values, as shown in \rfig{AOP}c for the distribution of $k$-resolved AOP dipole at Fermi level in a 2D constant-$k_x$ plane of Brillouin zone that contains low-energy Weyl nodes.

Finally, we plot the calculated nonlinear conductivity versus temperature curve in the inset of \rfig{RT}b, which shows a very good agreement with the experimental result. Because of uncertainties in the structure and magnetic configuration, we focus on the temperature range below 80~K (See the Supplemental Materials~\cite{SM_mNLHE}). First of all, the theoretical result gets the correct (negative) sign. Second, it reproduces the trend, namely, the rapid decrease of $\chi$ with temperature. Physically, this is mainly due to the temperature broadening of the Fermi surface, causing a cancellation of opposite contributions across small local gaps. Third, the order of magnitude of the calculated conductivity agrees with the experiment. For example, the experimental result of $\chi_{yxy}$ at 20 (50)~K is about $-4.1$~$(-1.8)$~$\Omega^{-1} \text{cm}^{-1} \text{T}^{-1}$, and the calculation gives $-8.7$~$(-3.5)$~$\Omega^{-1} \text{cm}^{-1} \text{T}^{-1}$.

Our results demonstrate the orbital mNLHE for the first time. The effect, complementary to the eNLHE, provides a probe for AOP, an intriguing band geometric quantity encoding the microscopic orbital magnetoelectric coupling. Furthermore, the giant orbital tunability of quantum geometry of electron wave functions can be expected in many quantum materials, especially topological semimetals, suggesting a new route for characterizing and controlling topological band features.

\acknowledgements
This work was supported by National Key Basic Research Program of China (Projects No. 2022YFA1403700 and No. 2020YFA0308800) and NSFC (Projects No. 12074009 and No. 11774009), by Singapore NRF CRP22-2019-0061, by the UGC/RGC of Hong Kong SAR (AoE/P-701/20), and by the Start-up Research Grant of University of Macau. We acknowledge computational support from National Supercomputing Centre Singapore and Texas Advanced Computing Center. Z.W. and Y.Y. were supported by National Key Basic Research Program of China (Project No. 2022YFA1403400), NSFC (Grant No. 92065109), and the Beijing Natural Science Foundation (Grants No. Z210006 and No. Z190006). Z.W. also thanks the Analysis and Testing Center at BIT for assistance in facility support.

%

\clearpage

\end{document}